\documentclass[10pt]{article}

\usepackage[english]{babel}

\usepackage{graphics,color}
\usepackage{newlfont,epsfig}

\begin{document}
\date{}
\begin{center}
{\Large\bf Two coupled qubits interacting with a thermal bath: A comparative study of 
different models}
\end{center}
\begin{center}
{\normalsize G.L. De\c cordi and A. Vidiella-Barranco \footnote{vidiella@ifi.unicamp.br}}
\end{center}
\begin{center}
{\normalsize{ Instituto de F\'\i sica ``Gleb Wataghin'' - Universidade Estadual de Campinas}}\\
{\normalsize{ 13083-859   Campinas  SP  Brazil}}\\
\end{center}
\begin{abstract}
We investigate the dynamics of two interacting two-level systems (qubits) having one of them isolated and the other coupled to  
a large number of modes of the quantized electromagnetic field (thermal reservoir). We consider two different models of 
system-reservoir interaction: i) a ``microscopic" model, according to 
which the corresponding master equation is derived taking into account the interaction between the two subsystems (qubits);
ii) a naive ``phenomenological" model, in which such interaction is neglected in the derivation of the master equation. 
We study the dynamics of quantities such as bipartite entanglement, quantum discord 
and the linear entropy of the isolated qubit in both the strong and weak coupling regimes of the inter-qubit interaction. 
We also consider different temperatures of the reservoir. We find significant disagreements between the results 
obtained from the two models even in the weak coupling regime. For instance, we show that according to the 
phenomenological model, the isolated qubit would approach a maximally mixed state more slowly for higher 
temperatures (unphysical result), while the microscopic model predicts the opposite behaviour (correct result).
\end{abstract}
The investigation of the coherent interaction between quantum subsystems is of fundamental importance in the 
field of quantum information processing. As quantum systems are normally susceptible to their environment, quantum behaviour 
may be substantially affected by unwanted couplings to their surroundings. It is therefore of relevance to be able 
to describe the environmental influence as accurately as possible. Several methods have been developed in order to treat 
such non-ideal quantum systems; for instance, models \cite{mazur65,loui65} involving the coupling of a system of 
interest with a thermal reservoir, e.g., a large number of modes of the electromagnetic field, may account for phenomena such as 
the irreversible loss of quantum coherence (decoherence) \cite{cald85,milb85}. 
Besides, the concept of decoherence is of central importance to the field of quantum information \cite{shor98}
as well as for the understanding of the emergence of the ``classical" world \cite{zeh70}. An example of practical 
(perturbative) approach is the one based on master equations for the reduced density operator \cite{loui65,milb85}, 
being largely employed to describe the dynamics of quantum systems weakly coupled to reservoirs. 
The system of interest may be constitued by a single or multiple quantum subsystems. We have observed that 
simple {\it ad hoc} models have been routinely used in the investigation of the dynamics of quantum coupled systems 
(such as interacting two-level systems) under the action of an environment. In such phenomenological models, 
the dissipative term of the master equation is normally derived by assuming the coupling of a thermal 
reservoir to a single sub-system. 

Notwithstanding, derivations of master equations which include the subsystems' interactions have appeared in the literature since the 
early seventies \cite{walls70,picard74,shib79}. Moreover, recent studies \cite{cress92,ron03,mess07,plenio10,mess11,semi14,neil14} 
show that phenomenological approaches may fail to give a proper evolution of the system's density operator. In our opinion, though, there is 
still a lack of works devoted to the discussion of the validity of system-reservoir interaction models describing composite quantum systems 
under the influence of a finite temperature bath. 

In this contribution we are going to consider two interacting two-level systems, being one of them (qubit 1) isolated from its
environment, and the other (qubit 2) in contact with a large number of modes of the quantized field (thermal reservoir). 
We will not make the rotating wave approximation 
for the qubit-qubit interaction, so that we are able to investigate the behaviour of the system in both the weak and strong 
coupling regimes. As a first step we will derive (closely following references \cite{sanchez11,breu02}) a microscopic master 
equation for our two-qubit system, by taking into account the interaction between them. We will then compare the results with the 
ones obtained from a naive phenomenological model, according to which the dissipative term of the master equation is derived 
by assuming the existence of a sole sub-system (qubit 2). Calculations are performed for a wide range of values of qubit-qubit coupling 
constants as well as for different temperatures of the reservoir. We also study this model from a different point of view; we consider 
qubit 1 as being coupled to a ``composite reservoir" constituted by the thermal bath plus qubit 2, i.e., we perform the trace 
over the qubit 2 variables and focus on the evolution of qubit 1. Our paper is organized as follows: in Section (2) we present the 
derivation of the microscopic master equation and its analytical solution. In Section (3) we address the strong coupling regime 
for the qubit-qubit interaction; we discuss several features of the solutions, such as the steady state 
of the two-qubit system, as well as the evolution of the bipartite (qubit-qubit) entanglement, quantum discord and the linear 
entropy associated to qubit 1. In Section (4) we present a study of the system in the weak coupling regime. In Section (5) we 
summarize our conclusions.

\section{Microscopic master equation for the two-qubit system}

\subsection{Interacting qubits: unitary evolution}

Our system of interest consists of two (dipole) coupled two-level systems (qubits 1 and 2), whose dynamics, without making the 
rotating wave approximation, is governed by the following Hamiltonian (in units of $\hbar$)

\begin{equation}
H_{S}=\Omega_{1}\sigma_{+}^{(1)}\sigma_{-}^{(1)}+\Omega_{2}\sigma_{+}^{(2)}\sigma_{-}^{(2)}
+\frac{\lambda}{2}\left(\sigma_{+}^{(1)}\sigma_{-}^{(2)}+\sigma_{-}^{(1)}\sigma_{+}^{(2)}
+\sigma_{+}^{(1)}\sigma_{+}^{(2)}+\sigma_{-}^{(1)}\sigma_{-}^{(2)}\right),\label{hamiltwoqubit}
\end{equation}
where $\sigma_{+}^{(i)}=\left|1^{(i)}\right\rangle \left\langle 0^{(i)}\right|$
and $\sigma_{-}^{(i)}=\left|0^{(i)}\right\rangle \left\langle 1^{(i)}\right|$
$(\mbox{with} \; i=1,\:2)$ are the raising and lowering operators for qubit 1 and 2,
respectively. Here $\Omega_i$ is the frequency of the $i-th$ qubit and $\lambda/2$ is the
coupling constant between the two qubits. The Hamiltonian above may be diagonalized in the 
uncoupled basis $\left\{ \left|0,0\right\rangle ;\left|1,0\right\rangle 
;\left|0,1\right\rangle ;\left|1,1\right\rangle \right\}$, with
eigenenergies and eigenstates (dressed states), in the resonant case, 
$\Omega_{1}=\Omega_{2}=\Omega$, given by

\begin{equation}
\begin{array}{ccccc}
E_{a}=\left(\Omega-\frac{\sqrt{\lambda^{2}+4\Omega^{2}}}{2}\right) &  &  &  & \left|a\right\rangle =\alpha_{+}\left|0,0\right\rangle -\alpha_{-}\left|1,1\right\rangle \\
\\
E_{b}=\left(\Omega-\frac{\lambda}{2}\right) &  &  &  & \left|b\right\rangle =\frac{1}{\sqrt{2}}\left|1,0\right\rangle -\frac{1}{\sqrt{2}}\left|0,1\right\rangle \\
\\
E_{c}=\left(\Omega+\frac{\lambda}{2}\right) &  &  &  & \left|c\right\rangle =\frac{1}{\sqrt{2}}\left|1,0\right\rangle +\frac{1}{\sqrt{2}}\left|0,1\right\rangle \\
\\
E_{d}=\left(\Omega+\frac{\sqrt{\lambda^{2}+4\Omega^{2}}}{2}\right) &  &  &  & \left|d\right\rangle =\alpha_{-}\left|0,0\right\rangle +\alpha_{+}\left|1,1\right\rangle,
\end{array}\label{eq:est energ contra}
\end{equation} 
with $\alpha_{\pm}=\sqrt{\frac{1}{2}\pm\frac{\Omega}{\sqrt{\lambda^{2}+4\Omega^{2}}}}.$ 

\subsection{Derivation of the microscopic master equation\label{derivmicro}}

Now we assume that qubit 1 is isolated from its environment (although it is coupled to qubit 2) 
and that qubit 2 is in contact with a thermal bath at temperature $T$. The bath itself consists of independent 
modes of the quantized electromagnetic field with Hamiltonian 
\begin{equation}
H_{B}=\sum_{n}\omega_{n}a_{n}^{\dagger}a_{n}.
\end{equation}
We consider the qubit 2-reservoir interaction as being dissipative, with effective interaction Hamiltonian of the form
\begin{equation}
H_{int}=\sigma_{x}^{\left(2\right)}\otimes B,\label{eq:inter q-r}
\end{equation}
where $B$ is the bath operator $B = \sum_{n}\varepsilon_{n}\left(a_{n}+a_{n}^{\dagger}\right)$,
$a_{n}^{\dagger}$ and $a_{n}$ are the creation and annihilation operators of photons corresponding to the $n-th$ mode of the field 
(frequency $\omega_{n}$), $\sigma_{x}^{\left(2\right)}=\sigma_{+}^{\left(2\right)}+\sigma_{-}^{\left(2\right)}$ 
(relative to qubit 2), and $\varepsilon_{n}$ is the coupling constant of qubit 2 to the $n-th$ mode of the field.
The total Hamiltionian, system of qubits plus bath is then $H  =  H_{S} + H_{B} + H_{int}$.

The master equation for the density operator, $\rho$, of the two qubit system
in the Born-Markov and rotating wave approximations is

\begin{equation}
\dot{\rho}\left(t\right)=-i\left[H_{S},\rho\left(t\right)\right]+\mathcal{D}\left(\rho\left(t\right)\right).
\end{equation}
The dissipative term may be written as \cite{breu02}
\begin{equation}
\mathcal{D}\left(\rho\left(t\right)\right)=\sum_{\omega}\gamma\left(\omega\right)\left(A\left(\omega\right)\rho\left(t\right)A^{\dagger}
\left(\omega\right)-\frac{1}{2}\left\{ A^{\dagger}\left(\omega\right)A\left(\omega\right),\rho\left(t\right)\right\} \right)\;.
\label{eq:termodiss}
\end{equation}
The rates $\gamma$ are given by
$\gamma\left(\omega\right)=\intop_{-\infty}^{+\infty}d\tau e^{i\omega\tau}\left\langle B^{\dagger}\left(\tau\right)B\left(0\right)\right\rangle$,
where $B(\tau)$ is the bath operator in the interaction representation, or
$B\left(\tau\right)=e^{i\, H_{B}\tau}B\: e^{-i\, H_{B}\tau}=\sum_{n}\varepsilon_{n}\left(a_{n}e^{-i\,\omega_{n}\tau}+a_{n}^{\dagger}e^{+i\,\omega_{n}\tau}\right)$,
and $\left\langle B^{\dagger}\left(\tau\right)B\left(0\right)\right\rangle \equiv \mbox{Tr}_{B}\left[B^{\dagger}\left(\tau\right)B\left(0\right)\rho_{B}\right]$
is the trace over variables of the field. Here
\begin{equation}
\rho_{B}=\frac{\exp\left(- H_{B}/kT\right)}{\mbox{Tr}_{B}\left\{ \exp\left(- H_{B}/kT\right)\right\} }\label{eq:distrcan}
\end{equation}
is the thermal state for the field bath at temperature $T$.
The jump operators $A(\omega)$ are defined as 
$A\left(\omega\right)\equiv\sum_{\epsilon^{\prime}-\epsilon=\omega}\Pi\left(\epsilon\right)A\,\Pi\left(\epsilon^{\prime}\right)$,
where $\Pi\left(\epsilon\right)$ is the projector acting on the sub-space associated to the energy eigenvalues $\epsilon$ of the 
Hamiltonian $H_{S}$, and the summation is over the eigenstates having fixed energy difference equal to $\omega$ (in units of $\hbar$). 
In our case, $A = \sigma_{x}^{\left(2\right)}$.
The first Bohr frequency is
\[
\omega_{I}=\left(\sqrt{\lambda^{2}+4\Omega^{2}}-\lambda\right)/2
\]
for the transitions $\left|b\right\rangle \rightarrow\left|a\right\rangle $
and $\left|d\right\rangle \rightarrow\left|c\right\rangle $ and is related
to the jump operator
\begin{equation}
\sigma_{x}^{\left(2\right)}\left(\omega_{I}\right)=\left\langle a\right|\sigma_{x}^{\left(2\right)}
\left|b\right\rangle \left|a\right\rangle \left\langle b\right|+\left\langle c\right|
\sigma_{x}^{\left(2\right)}\left|d\right\rangle \left|c\right\rangle \left\langle d\right|\;,\label{eq:salto I}
\end{equation}
while the second Bohr frequency
\[
\omega_{II}=\left(\sqrt{\lambda^{2}+4\Omega^{2}}+\lambda\right)/2
\]
for the transitions $\left|c\right\rangle \rightarrow\left|a\right\rangle $
and $\left|d\right\rangle \rightarrow\left|b\right\rangle $ is related to
\begin{equation}
\sigma_{x}^{\left(2\right)}\left(\omega_{II}\right)=\left\langle a\right|\sigma_{x}^{\left(2\right)}
\left|c\right\rangle \left|a\right\rangle \left\langle c\right|+\left\langle b\right|
\sigma_{x}^{\left(2\right)}\left|d\right\rangle \left|b\right\rangle \left\langle d\right|\;.\label{eq:salto II}
\end{equation}

After identifying each term in Eq. (\ref{eq:termodiss}), we may rewrite it as
\begin{eqnarray}
\mathcal{D}\left(\rho\left(t\right)\right) & = & \sum_{i=I}^{II}\gamma\left(\omega_{i}\right)\left(\sigma_{x}^{\left(2\right)}\left(\omega_{i}\right)\rho\left(t\right)\sigma_{x}^{\left(2\right)\,\dagger}\left(\omega_{i}\right)-\frac{1}{2}\left\{ \sigma_{x}^{\left(2\right)\,\dagger}\left(\omega_{i}\right)\sigma_{x}^{\left(2\right)}\left(\omega_{i}\right),\rho\left(t\right)\right\} \right) \nonumber\\
 &  & +\sum_{i=I}^{II}\overline{\gamma}\left(\omega_{i}\right)\left(\sigma_{x}^{\left(2\right)\,\dagger}\left(\omega_{i}\right)\rho\left(t\right)\sigma_{x}^{\left(2\right)}\left(\omega_{i}\right)-\frac{1}{2}\left\{ \sigma_{x}^{\left(2\right)}\left(\omega_{i}\right)\sigma_{x}^{\left(2\right)\,\dagger}\left(\omega_{i}\right),\rho\left(t\right)\right\} \right)\;,\nonumber \\
\end{eqnarray}
with the Kubo-Martin-Schwinger relation \cite{breu02} 
\begin{equation}
\overline{\gamma}\left(\omega_{i}\right)=\exp\left(-\omega_{i}/kT\right)\gamma\left(\omega_{i}\right),\;\label{gammabar}
\end{equation}
and $\;\sigma_{x}^{\left(2\right)\,\dagger}\left(\omega_{i}\right)=\sigma_{x}^{\left(2\right)}\left(-\omega_{i}\right)$.

Now, working out the master equation above using the expressions (\ref{eq:est energ contra}) for the eigenstates of $H_{S}$ 
as well as the jump operators (\ref{eq:salto I}) and (\ref{eq:salto II}), the microscopic master equation will finally read
\begin{eqnarray}
\dot{\rho}\left(t\right) & = & -i\left[H_{S},\rho\left(t\right)\right]\\
&  &+\:c_{I}\left(\left|a\right\rangle \left\langle b\right|\rho\left|b\right\rangle \left\langle a\right|-\frac{1}{2}\left\{ \left|b\right\rangle \left\langle b\right|,\rho\right\} \right)+c_{II}\left(\left|a\right\rangle \left\langle c\right|\rho\left|c\right\rangle \left\langle a\right|-\frac{1}{2}\left\{ \left|c\right\rangle \left\langle c\right|,\rho\right\} \right)\nonumber \\
&  & +\: c_{II}\left(\left|b\right\rangle \left\langle d\right|\rho\left|d\right\rangle \left\langle b\right|-\frac{1}{2}\left\{ \left|d\right\rangle \left\langle d\right|,\rho\right\} \right)+c_{I}\left(\left|c\right\rangle \left\langle d\right|\rho\left|d\right\rangle \left\langle c\right|-\frac{1}{2}\left\{ \left|d\right\rangle \left\langle d\right|,\rho\right\} \right)\nonumber \\
&  & +\:\overline{c}_{I}\left(\left|b\right\rangle \left\langle a\right|\rho\left|a\right\rangle \left\langle b\right|-\frac{1}{2}\left\{ \left|a\right\rangle \left\langle a\right|,\rho\right\} \right)+\overline{c}_{I}\left(\left|d\right\rangle \left\langle c\right|\rho\left|c\right\rangle \left\langle d\right|-\frac{1}{2}\left\{ \left|c\right\rangle \left\langle c\right|,\rho\right\} \right)\nonumber \\
&  & +\:\overline{c}_{II}\left(\left|d\right\rangle \left\langle b\right|\rho\left|b\right\rangle \left\langle d\right|-\frac{1}{2}\left\{ \left|b\right\rangle \left\langle b\right|,\rho\right\} \right)+\overline{c}_{II}\left(\left|c\right\rangle \left\langle a\right|\rho\left|a\right\rangle \left\langle c\right|-\frac{1}{2}\left\{ \left|a\right\rangle \left\langle a\right|,\rho\right\} \right)\nonumber \\
&  & -\: c_{I}\left(\left|a\right\rangle \left\langle b\right|\rho\left|d\right\rangle \left\langle c\right|+\left|c\right\rangle \left\langle d\right|\rho\left|b\right\rangle \left\langle a\right|\right)+c_{II}\left(\left|a\right\rangle \left\langle c\right|\rho\left|d\right\rangle \left\langle b\right|+\left|b\right\rangle \left\langle d\right|\rho\left|c\right\rangle \left\langle a\right|\right)\nonumber \\
&  & -\:\overline{c}_{I}\left(\left|b\right\rangle \left\langle a\right|\rho\left|c\right\rangle \left\langle d\right|+\left|d\right\rangle \left\langle c\right|\rho\left|a\right\rangle \left\langle b\right|\right)+\overline{c}_{II}\left(\left|d\right\rangle \left\langle b\right|\rho\left|a\right\rangle \left\langle c\right|+\left|c\right\rangle \left\langle a\right|\rho\left|b\right\rangle \left\langle d\right|\right),\nonumber\label{mestramic}
\end{eqnarray}
where the decay constants are given by

\begin{eqnarray}
c_{I}=\alpha^{2}\gamma\left(\omega_{I}\right) &  & c_{II}=\eta^{2}\gamma\left(\omega_{II}\right) \nonumber
\\
\overline{c}_{I}=\alpha^{2}\overline{\gamma}\left(\omega_{I}\right) &  & \overline{c}_{II}=\eta^{2}\overline{\gamma}\left(\omega_{II}\right)\;,
\label{coeffs}
\end{eqnarray}
with 

\begin{eqnarray}
\alpha&=&-\frac{1}{2}\left(\sqrt{1+\frac{2\Omega}{\sqrt{\lambda^{2}+4\Omega^{2}}}}+\sqrt{1-\frac{2\Omega}{\sqrt{\lambda^{2}+4\Omega^{2}}}}\right)\,,\\
\nonumber \\
\qquad\eta&=&\frac{1}{2}\left(\sqrt{1+\frac{2\Omega}{\sqrt{\lambda^{2}+4\Omega^{2}}}}-\sqrt{1-\frac{2\Omega}{\sqrt{\lambda^{2}+4\Omega^{2}}}}\right)\,,
\end{eqnarray}
and $\overline{\gamma}\left(\omega_{i}\right)=\gamma\left(\omega_{i}\right)e^{-\beta\omega_{i}}$.

The functions $\gamma\left(\omega_{i}\right)$ are related (see, for instance \cite{breu02}) to the spectral density of the reservoir ($J\left(\omega_{i}\right)$) through

\begin{equation}
\gamma\left(\omega_{i}\right)=J\left(\omega_{i}\right)\left[1+\overline{n}\left(\omega_{i}\right)\right]\:,\label{eq:gama micro}
\end{equation}
where the index $i = I,II$ corresponds to the Bohr frequencies of the model.
Here $\overline{n}\left(\omega_{i}\right)$ is the mean photon number associated to
the mode of frequency $\omega_{i}$ of a thermal state at temperature $T$, 
\begin{equation}
\overline{n}\left(\omega_{i}\right)=\frac{1}{e^{\omega_{i}/kT}-1}.\label{nmedio}
\end{equation}
We have chosen a Lorentzian function for the spectral density, or%
\begin{equation}
J\left(\omega_{i}\right)=\frac{\gamma_{0}\Gamma^{2}}{\left(\omega_{i}-\Omega_{0}\right)^{2}+\Gamma^{2}}\,,\label{spectral}
\end{equation}
where $\gamma_{0}$ is the single qubit decay rate, $\Omega_0$ is the central frequency, and $\Gamma$ is the half-width of 
the distribution. 

The sets of coupled differential equations for the matrix elements of the two qubit density operator 
in the microscopic model, as well as their analytical solutions, may be found in Appendix A.

\subsection{Phenomenological master equation}

We now consider the phenomenological approach, according to which the reservoir is coupled to a single qubit (qubit 2) neglecting the fact that
such a system is in interaction with a second qubit (qubit 1). We assume the same (qubit 2-reservoir) interaction Hamiltonian as employed in the microscopic case 
[see Eq. (\ref{eq:inter q-r})] and follow the procedure developed in Section \ref{derivmicro}. However, in order to build the Lindblad operators, we use here the 
(one qubit) bare states rather than the (two qubit) dressed states employed in the microscopic master equation, or

\begin{equation}
A(\Omega) = |0\rangle\langle 0| \sigma_{x}^{\left(2\right)} |1\rangle\langle 1| = |0\rangle\langle 0| \left(\sigma_{+}^{\left(2\right)}
+ \sigma_{-}^{\left(2\right)} \right)|1\rangle\langle 1| = 
\sigma_{-}^{\left(2\right)}.
\end{equation}
Analogously, we obtain $A^\dagger(\Omega) = \sigma_{+}^{\left(2\right)}$.

Thus, the phenomenological master equation will read

\begin{eqnarray}
\dot{\rho}\left(t\right) & = & -i\left[H_{S},\rho\left(t\right)\right]+\gamma\left(\sigma_{-}^{\left(2\right)}\rho\left(t\right)\sigma_{+}^{\left(2\right)}
-\frac{1}{2}\left\{ \sigma_{+}^{\left(2\right)}\sigma_{-}^{\left(2\right)},\rho\left(t\right)\right\} \right)\nonumber \\
 &  & +\,\overline{\gamma}\left(\sigma_{+}^{\left(2\right)}\rho\left(t\right)\sigma_{-}^{\left(2\right)}
-\frac{1}{2}\left\{ \sigma_{-}^{\left(2\right)}\sigma_{+}^{\left(2\right)},\rho\left(t\right)\right\} \right)\:,\label{eq:mestra fenom}
\end{eqnarray}
where $H_S$ is the two-qubit Hamiltonian in Eq. (\ref{hamiltwoqubit}).
Here, similarly to the microscopic case, the function $\gamma\equiv\gamma\left(\Omega\right)$ and the spectral density
$J\left(\Omega\right)$ are related through
$\gamma\left(\Omega\right)=J\left(\Omega\right)\left[1+\overline{n}\left(\Omega\right)\right]$.
The quantities $\overline{\gamma}\equiv\overline{\gamma}\left(\Omega\right)$, $\overline{n}\left(\Omega\right)$ and 
$J\left(\Omega\right)$ are just the same expressions as in Eqs. (\ref{gammabar}), (\ref{nmedio}) and (\ref{spectral}) 
respectively, but having $\Omega$ as argument, instead. In order to make a trustworthy comparison between the two models, 
we have assured that $J(\omega_i)\approx J(\Omega)$, with $\overline{n}(\omega_i)\approx\overline{n}(\Omega)$.  

The resulting differential equations for the relevant matrix elements of the two qubit density operator 
in the phenomenological model may be found in Appendix B.
In this case the differential equations have been numerically solved. 

\section{Comparison between microscopic and phenomenological models: strong coupling regime}

Now we consider the strong coupling regime for the qubit-qubit interaction, or $\lambda \geq \Omega$.
Before addressing some general features of the system evolution, we would like to discuss the behaviour 
of the two-qubit steady state. In order to compare the predictions obtained from the microscopic and phenomenological 
models, we first calculate explicitly the stationary state of the reduced density operator for the two qubit system. 

\subsection{Steady state analysis}

\subsubsection{Microscopic model}

In the microscopic approach, the steady state is given by 

\[
\rho_{\infty,m}=\rho_{aa}\left|a\right\rangle \left\langle a\right|+\rho_{bb}\left|b\right\rangle \left\langle b\right|
+\rho_{cc}\left|c\right\rangle \left\langle c\right|+\rho_{dd}\left|d\right\rangle \left\langle d\right|\,,
\]
with the elements

\[
\rho_{aa}=\frac{c_I c_{II}}{\left(c_{I}+\bar{c}_{I}\right)\left(c_{II}+\bar{c}_{II}\right)},\qquad\rho_{bb}=\frac{\bar{c}_{I}
c_{II}}{\left(c_{I}+\bar{c}_{I}\right)\left(c_{I}+\bar{c}_{II}\right)},
\]

\[
\rho_{cc}=\frac{c_{I}\bar{c}_{II}}{\left(c_{I}+\bar{c}_{I}\right)\left(c_{II}+\bar{c}_{II}\right)},\qquad\rho_{dd}=
\frac{\bar{c}_{I}\bar{c}_{II}}{\left(c_{I}+\bar{c}_{I}\right)\left(c_{II}+\bar{c}_{II}\right)}\,.
\]

The coefficients $c_i(\overline{c}_i)$ are defined above, in the relations (\ref{coeffs}).
From the relations $\overline{c}_{I}$ and $\overline{c}_{II}$ it is possible to show that $\rho_{aa}$ may be written as
\[
\rho_{aa}=\frac{1}{1+e^{-\beta\omega_{I}}+e^{-\beta\omega_{II}}+e^{-\beta\left(\omega_{I}+\omega_{II}\right)}}\,,
\]
where $\omega_{I}=\left(\sqrt{\lambda^{2}+4\Omega^{2}}-\lambda\right)/2$
and $\omega_{II}=\left(\sqrt{\lambda^{2}+4\Omega^{2}}+\lambda\right)/2$
are the Bohr frequencies, which are related to the relevant energy differences,
$\omega_{I}=E_{b}-E_{a},\qquad\omega_{II}=E_{c}-E_{a},\qquad\omega_{I}+\omega_{II}=E_{d}-E_{a}$.
Thus we obtain the following expression for the matrix element $\rho_{aa}$ 

\begin{eqnarray}
\rho_{aa} & = & \frac{1}{1+e^{-\beta\left(E_{b}-E_{a}\right)}+e^{-\beta\left(E_{c}-E_{a}\right)}+e^{-\beta\left(E_{d}-E_{a}\right)}}\,,\nonumber \\
 & = & \frac{e^{-\beta E_{a}}}{e^{-\beta E_{a}}}\left(\frac{1}{1+e^{-\beta\left(E_{b}-E_{a}\right)}+e^{-\beta\left(E_{c}-E_{a}\right)}
+e^{-\beta\left(E_{d}-E_{a}\right)}}\right),\label{eq:elem principal}\\
 & = & \frac{e^{-\beta E_{a}}}{e^{-\beta E_{a}}+e^{-\beta E_{b}}+e^{-\beta E_{c}}+e^{-\beta E_{d}}}\,.\nonumber 
\end{eqnarray}

The other matrix elements may be obtained from $\rho_{aa}$, or
$\rho_{bb}=e^{-\beta\omega_{I}}\rho_{aa},
\:\rho_{cc}=e^{-\beta\omega_{II}}\rho_{aa},
\:\rho_{dd}=e^{-\beta\left(\omega_{I}+\omega_{II}\right)}\rho_{aa},$
which, together with (\ref{eq:elem principal}) gives
\begin{eqnarray}
\rho_{bb} & = & \frac{e^{-\beta E_{b}}}{e^{-\beta E_{a}}+e^{-\beta E_{b}}+e^{-\beta E_{c}}+e^{-\beta E_{d}}},\nonumber \\
\rho_{cc} & = & \frac{e^{-\beta E_{c}}}{e^{-\beta E_{a}}+e^{-\beta E_{b}}+e^{-\beta E_{c}}+e^{-\beta E_{d}}},\label{eq:restante contra-1}\\
\rho_{dd} & = & \frac{e^{-\beta E_{d}}}{e^{-\beta E_{a}}+e^{-\beta E_{b}}+e^{-\beta E_{c}}+e^{-\beta E_{d}}}\,.\nonumber 
\end{eqnarray}

Thus, the resulting steady state corresponds to a state which is in thermal equilibrium with the reservoir.

\subsubsection{Phenomenological model}

The density operator in the phenomenological approach is given by

\begin{eqnarray}
\rho_{\infty,p} & = & \rho_{11}\left|0,0\right\rangle \left\langle 0,0\right|+\rho_{22}\left|0,1\right\rangle \left\langle 0,1\right|
+\rho_{33}\left|1,0\right\rangle \left\langle 1,0\right|+\rho_{44}\left|1,1\right\rangle \left\langle 1,1\right|\nonumber \\
&  & +\rho_{23}\left|0,1\right\rangle \left\langle 1,0\right|
+\rho_{23}^{*}\left|1,0\right\rangle \left\langle 0,1\right|+\rho_{14}\left|0,0\right\rangle \left\langle 1,1\right|
+\rho_{14}^{*}\left|1,1\right\rangle \left\langle 0,0\right|\,,\nonumber\label{eq:op fen}\\
\end{eqnarray}
with populations and coherences

\[
\rho_{11}=\frac{\left(3\gamma^{3}\bar{\gamma}+\gamma^{2}\left(3\bar{\gamma}^{2}+\lambda^{2}+16\Omega^{2}\right)+\gamma\left(2\lambda^{2}\bar{\gamma}
+\bar{\gamma}^{3}\right)+\lambda^{2}\bar{\gamma}^{2}+\gamma^{4}\right)}{2\left(\bar{\gamma}+\gamma\right)^{2}\left(\left(\bar{\gamma}
+\gamma\right)^{2}+2\lambda^{2}+8\Omega^{2}\right)}
\]

\[
\rho_{22}=\frac{\left(\gamma^{3}\bar{\gamma}+\gamma^{2}\left(3\bar{\gamma}^{2}+\lambda^{2}\right)+\gamma\bar{\gamma}\left(3\bar{\gamma}^{2}
+2\left(\lambda^{2}+8\Omega^{2}\right)\right)+\bar{\gamma}^{2}\left(\bar{\gamma}^{2}+\lambda^{2}\right)\right)}{2\left(\bar{\gamma}
+\gamma\right)^{2}\left(\left(\bar{\gamma}+\gamma\right)^{2}+2\lambda^{2}+8\Omega^{2}\right)}
\]

\[
\rho_{33}=\frac{\left(3\gamma^{3}\bar{\gamma}+\gamma^{2}\left(3\bar{\gamma}^{2}+\lambda^{2}\right)+\gamma\bar{\gamma}\left(\bar{\gamma}^{2}
+2\left(\lambda^{2}+8\Omega^{2}\right)\right)+\lambda^{2}\bar{\gamma}^{2}+\gamma^{4}\right)}{2\left(\bar{\gamma}
+\gamma\right)^{2}\left(\left(\bar{\gamma}+\gamma\right)^{2}+2\lambda^{2}+8\Omega^{2}\right)}
\]

\[
\rho_{44}=\frac{\left(\gamma^{3}\bar{\gamma}+\gamma^{2}\left(3\bar{\gamma}^{2}+\lambda^{2}\right)+\bar{\gamma}^{2}\left(\bar{\gamma}^{2}
+\lambda^{2}+16\Omega^{2}\right)+\gamma\left(2\lambda^{2}\bar{\gamma}+3\bar{\gamma}^{3}\right)\right)}{2\left(\bar{\gamma}
+\gamma\right)^{2}\left(\left(\bar{\gamma}+\gamma\right)^{2}+2\lambda^{2}+8\Omega^{2}\right)}
\]

\[
\rho_{23}=\frac{i\,\lambda\left(\bar{\gamma}-\gamma\right)}{2\left(\left(\bar{\gamma}+\gamma\right)^{2}+2\lambda^{2}+8\Omega^{2}\right)}
\]

\[
\rho_{14}=\frac{2\lambda\Omega\left(\gamma-\bar{\gamma}\right)}{\left(\bar{\gamma}+\gamma\right)\left(\left(\bar{\gamma}+\gamma\right)^{2}
+2\lambda^{2}+8\Omega^{2}\right)}+i\,\frac{\lambda\left(\gamma-\bar{\gamma}\right)}{2\left(\left(\bar{\gamma}+\gamma\right)^{2}+2\lambda^{2}+8\Omega^{2}\right)}\,.
\]

The density operator in (\ref{eq:op fen}) may be rewritten using

\[
\left|0,0\right\rangle =\alpha_{+}\left|a\right\rangle +\alpha_{-}\left|d\right\rangle ,\quad\left|0,1\right\rangle =\frac{\left|c\right\rangle -\left|b\right\rangle }{\sqrt{2}},
\]

\[
\left|1,1\right\rangle =\alpha_{+}\left|d\right\rangle -\alpha_{-}\left|a\right\rangle ,\quad\left|1,0\right\rangle =\frac{\left|c\right\rangle +\left|b\right\rangle }{\sqrt{2}},
\]
where $\alpha_{\pm}=\sqrt{\frac{1}{2}\pm\frac{\Omega}{\sqrt{\lambda^{2}+4\Omega^{2}}}}$.

Thus, in the dressed state basis the phenomenological steady state density operator reads 

\begin{eqnarray}
\rho_{\infty,p} & = & \left(\alpha_{-}^{2}\rho_{44}+\alpha_{+}^{2}\rho_{11}-2\alpha_{+}\alpha_{-}Re\,[\rho_{14}]\right)\left|a\right\rangle \left\langle a\right|
+\left(\frac{\rho_{22}}{2}+\frac{\rho_{33}}{2}-Re\,[\rho_{23}]\right)\left|b\right\rangle \left\langle b\right|\nonumber \\ 
 &  & +\left(\frac{\rho_{22}}{2}+\frac{\rho_{33}}{2}+Re\,[\rho_{23}]\right)\left|c\right\rangle \left\langle c\right|+\left(\alpha_{-}^{2}\rho_{11}
+\alpha_{+}^{2}\rho_{44}+2\alpha_{+}\alpha_{-}Re\,[\rho_{14}]\right)\left|d\right\rangle \left\langle d\right|\nonumber \\ 
\label{steadyphen}\\
 &  & +\left(\alpha_{+}\alpha_{-}\left(\rho_{11}-\rho_{44}\right)+\alpha_{+}^{2}\rho_{14}-\alpha_{-}^{2}\rho_{14}^{*}\right)\left|a\right\rangle \left\langle d\right|
+\left(\frac{\rho_{33}}{2}-\frac{\rho_{22}}{2}-i\, Im\,[\rho_{23}]\right)\left|b\right\rangle \left\langle c\right|\nonumber \\
 &  & +\left(\alpha_{+}\alpha_{-}\left(\rho_{11}-\rho_{44}\right)+\alpha_{+}^{2}\rho_{14}^{*}-\alpha_{-}^{2}\rho_{14}\right)\left|d\right\rangle \left\langle a\right|
+\left(\frac{\rho_{33}}{2}-\frac{\rho_{22}}{2}+i\, Im\,[\rho_{23}]\right)\left|c\right\rangle \left\langle b\right|\,,
\nonumber 
\end{eqnarray}
which is certainly not a thermal equilibrium state. 
Therefore our results are a clear indication of the inadequacy of the phenomenological model when applied to a two qubit system
in contact with a thermal reservoir. This is in accord to the discussions found in references \cite{cress92,mess07}, in which
similar conclusions are drawn from the analysis of the dissipative Jaynes-Cummings model.

We consider now the thermal bath in its (multimode) vacuum state, i.e., $T = 0$ K. In this specific situation, energy is irreversibly transferred from the two-qubit 
system to the reservoir, and we expect the system to relax to its minimum energy state. As we have already seen, the ground state of the two qubit system is
$\left|\phi\right\rangle =\alpha_{+}\left|0,0\right\rangle -\alpha_{-}\left|1,1\right\rangle$, which is in general 
an entangled state. Note that the state $\left|\phi\right\rangle$ becomes a product state only in the limit of very weak coupling, 
or $\lambda \ll \Omega$, for which $\alpha_- \rightarrow 0$. However in this section we are considering the strong coupling regime 
($\lambda \geq \Omega$), instead. The predictions of the microscopic and phenomenological models are very different in this case: 
according to the microscopic model, the asymptotic two-qubit density operator is
\begin{eqnarray}
\rho_{\infty,m} & = & \left|a\right\rangle \left\langle a\right| \nonumber \\ 
 & = & \left(\alpha_{+}\left|0,0\right\rangle -\alpha_{-}\left|1,1\right\rangle \right)
\left(\alpha_{+}\left\langle 0,0\right|-\alpha_{-}\left\langle 1,1\right|\right),\label{eq:asymptmic}
\end{eqnarray}
which coincides with the ground state of the system, while according to the phenomenological model we have the 
density operator in equation (\ref{steadyphen}) above, a very different state. This is another evidence of the inadequacy 
of the {\it ad hoc} model. 

\subsection{Degree of bipartite entanglement}
 
The concurrence \cite{woot97}, an entanglement monotone employed to quantify quantum entanglement, is defined as 
\begin{equation}
C\left(t\right)=max[0,\sqrt{\xi_{1}\left(t\right)}-\sqrt{\xi_{2}\left(t\right)}-\sqrt{\xi_{3}\left(t\right)}-
\sqrt{\xi_{4}\left(t\right)}]\,,\label{eq:concorrencia}
\end{equation}
where $\xi_{i}\left(t\right)$ are the eigenvalues of the matrix
$M\left(t\right)=\rho\left(t\right)\tilde{\rho}\left(t\right)$ placed in decreasing order in Eq. (\ref{eq:concorrencia}),
with $\tilde{\rho}\left(t\right)=\sigma_{y}\otimes\sigma_{y}\,\rho^{*}\left(t\right)\sigma_{y}\otimes\sigma_{y}$ and where $\sigma_{y}$ 
is the usual Pauli matrix. Now we are going to calculate the concurrence as a function of time, for different temperatures of the reservoir 
in both the microscopic and the phenomenological model. 

For the microscopic model, the concurrence is given by

\begin{equation}
C\left(t\right)=2\, max[0,\left|Q_{14}\left(t\right)\right|-\sqrt{Q_{22}\left(t\right)Q_{33}\left(t\right)},\left|Q_{23}\left(t\right)\right|-\sqrt{Q_{11}\left(t\right)Q_{44}\left(t\right)}]\,,\label{eq:conc microscopica}
\end{equation}
where

\[
\begin{array}{ccccc}
Q_{11}\left(t\right)=\alpha_{+}^{2}\rho_{aa}+\alpha_{-}^{2}\rho_{dd} &  &  &  & Q_{22}\left(t\right)=\frac{1}{2}\left(\rho_{bb}+\rho_{cc}\right)-Re\left(\rho_{bc}\right)\,,\\
\\
Q_{33}\left(t\right)=\frac{1}{2}\left(\rho_{bb}+\rho_{cc}\right)+Re\left(\rho_{bc}\right) &  &  &  & Q_{44}\left(t\right)=\alpha_{-}^{2}\rho_{aa}+\alpha_{+}^{2}\rho_{dd},\\
\\
Q_{14}\left(t\right)=\alpha_{+}\alpha_{-}\left(\rho_{dd}-\rho_{aa}\right) &  &  &  & Q_{23}\left(t\right)=\frac{1}{2}\left(\rho_{cc}-\rho_{bb}\right)-i\, Im\left(\rho_{bc}\right).\\
\\
\end{array}
\]

The corresponding concurrence in the phenomenological model is given by the same expression as equation (\ref{eq:conc microscopica}) 
but having the elements $Q_{ij}$ being replaced by the matrix elements themselves, $\rho_{ij}$. 

\subsection{Quantum discord}

More general quantum correlations may be described by the quantum discord \cite{zurek01}. The discord is defined in terms of the mutual information 
shared by two quantum subsystems (qubit 1 and qubit 2, for instance), or
\begin{equation}
I(\rho) = S(\rho_{q1}) + S(\rho_{q2}) - S(\rho)\,,\label{eq:mutualinf}
\end{equation}
where we have made $\rho_{q1q2} \equiv \rho$ and $S(\rho) = - \mbox{Tr}(\rho\log_2 \rho)$ is the von Neumann entropy. 
Consider a measurement over subsystem 
qubit 2, $\left\{\Pi_i\right\}$, with $p_i = \mbox{Tr}(\Pi_i \rho)$ being the probability of the $i$-th measurement outcome and 
$\rho^i_{q1} = \mbox{Tr}_{q2}(\Pi_i \rho)/p_i$ the post-measurement state. The classical correlations are defined as 
$J_2(\rho) = \max_{\left\{\Pi_i\right\}} J_{\left\{\Pi_i\right\}}(\rho)$,
where $J_{\left\{\Pi_i\right\}}(\rho) = S(\rho_{q1}) - \sum_i p_i S(\rho^i_{q1})$. The maximization is over all measurements and
$\rho_{q1} = \mbox{Tr}_{q2} \left[\rho\right]$.
The quantum discord ${\cal D}$ is defined as the difference between the mutual information $I(\rho)$,
which represents the total correlations, and the classical correlations $J_2(\rho)$, or
\begin{equation}
{\cal D}_2 (\rho) = I(\rho) - J_2(\rho) = S(\rho_{q2}) - S(\rho) + \min_{\left\{\Pi_i\right\}} \sum_i p_i S(\rho^i_{q1}) \,.\label{eq:discord}
\end{equation}
Note that we may have ${\cal D}_2 \neq {\cal D}_1$.

As the calculation of discord involves an optimization procedure, analytical results are rare. Nevertheless, in our case we may employ
an approximate expression for two qubit states \cite{james12}  
\begin{equation}
{\cal D}_2 (\rho) \approx S(\rho_{q2}) - S(\rho) + \min(N_1,N_2)\,,\label{eq:discordapprox}
\end{equation}
where the von Neumann entropies are
\begin{equation}
S\left(\rho_{q2}\right)=-\left(Q_{11}+Q_{33}\right)\log_{2}\left(Q_{11}+Q_{33}\right) - \left(Q_{22}+Q_{44}\right)\log_{2}\left(Q_{22}+Q_{44}\right)\,,\label{eq:entropy2}
\end{equation}
and
\begin{equation}
S\left(\rho\right)=-\sum_{j=1}^{4}\Lambda_{j}\log_{2}\Lambda_{j}\,,\label{eq:entropy x}
\end{equation}
with
\begin{eqnarray}
\Lambda_{1}&=&\frac{1}{2}\left[\left(Q_{11}+Q_{44}\right)+\sqrt{\left(Q_{11}-Q_{44}\right)^{2}+4\left|Q_{14}\right|^{2}}\right]\nonumber\\ 
\Lambda_{2}&=&\frac{1}{2}\left[\left(Q_{11}+Q_{44}\right)-\sqrt{\left(Q_{11}-Q_{44}\right)^{2}+4\left|Q_{14}\right|^{2}}\right]\,, \nonumber
\end{eqnarray}
\begin{eqnarray}
\Lambda_{3}&=&\frac{1}{2}\left[\left(Q_{22}+Q_{33}\right)+\sqrt{\left(Q_{22}-Q_{33}\right)^{2}+4\left|Q_{23}\right|^{2}}\right]\nonumber\\ 
\Lambda_{4}&=&\frac{1}{2}\left[\left(Q_{22}+Q_{33}\right)-\sqrt{\left(Q_{22}-Q_{33}\right)^{2}+4\left|Q_{23}\right|^{2}}\right]\,. \nonumber
\end{eqnarray}
Also
\begin{equation}
N_{1}=-y\, \log_{2}y-(1-y)\, \log_{2}(1-y)\,,\label{eq:binary}
\end{equation}
where

\begin{equation}
y=\frac{1+\sqrt{\left(Q_{11}-Q_{44}+Q_{22}-Q_{33}\right)^{2}+4\left(\left|Q_{14}\right|+\left|Q_{23}\right|\right)^{2}}}{2}\,,\label{eq:aux binary}
\end{equation}
and

\begin{eqnarray}
N_{2}&=&-Q_{11}\, \log_{2}\left(\frac{Q_{11}}{Q_{11}+Q_{33}}\right)-Q_{22}\, \log_{2}\left(\frac{Q_{22}}{Q_{22}+Q_{44}}\right)\nonumber\\
&-&Q_{33}\, \log_{2}\left(\frac{Q_{33}}{Q_{33}+Q_{11}}\right)-Q_{44}\, \log_{2}\left(\frac{Q_{44}}{Q_{44}+Q_{22}}\right)\,.\label{eq:n2}
\end{eqnarray}

\subsection{Linear entropy of qubit 1: composite reservoir}

We may regard the system considered here from a different perspective: while keeping exactly the same 
configuration, we may view qubit 1 as a single quantum subsystem coupled to a more complex, ``composite reservoir", 
constituted by qubit 2 plus the thermal bath. In other words, we trace over the qubit 2 variables and analyse the 
behaviour of the qubit 1 dynamics. The coherence properties of qubit 1 may be described by its linear entropy, defined as
$S\left(t\right)=1-\mbox{Tr}\left[\rho_{q1}^{2}\left(t\right)\right]$.
The linear entropy relative to qubit 1, according to the microscopic model, is
\[
S_m\left(t\right)=2P_{0}\left(t\right)\left[1-P_{0}\left(t\right)\right]\,,
\]
with $P_{0}\left(t\right)=Q_{11}\left(t\right)+Q_{22}\left(t\right)$. 

For the phenomenological model, the linear entropy is given by
\begin{equation}
S_p\left(t\right)=1-\left[\rho_{11}\left(t\right)+\rho_{22}\left(t\right)\right]^{2}
-\left[\rho_{33}\left(t\right)+\rho_{44}\left(t\right)\right]^{2}\label{eq:e fenom}.
\end{equation} 

\subsection{Numerical results: strong coupling regime}
 
\begin{figure}
	\centering
		\includegraphics{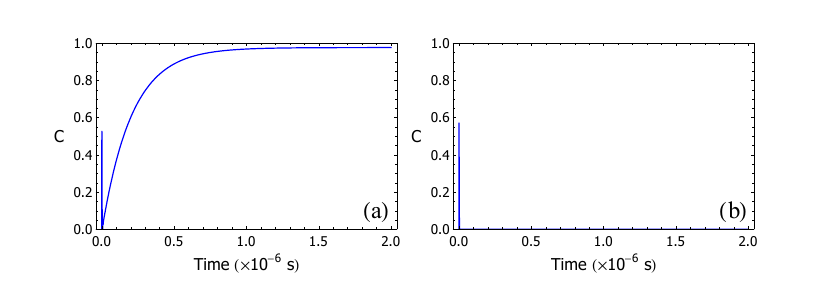}
	\caption{(Color online) Concurrence between qubits 1 and 2 as a function of time in the very strong coupling regime
	for an initial two qubit state $|\Psi\rangle_{q1,q2}=|1,0\rangle$ according to
	(a) the microscopic model; (b) the phenomenological model. 
	In both plots $\Omega = 4 \times 10^{8} s^{-1}$, $\lambda = 10\Omega$,
	$\gamma_0 = 0.01\times 5 \times 10^{10} s^{-1}$, $\Gamma = 5 \times 10^{10} s^{-1}$, 
	$\Omega_0 = 2\Omega$ and $T \approx 0$K.}
	\label{figure1}	
\end{figure}

Before showing some numerical results related to dynamics of entanglement, discord and the linear 
entropy, we would like to recall the discussion of the previous subsection, related to the steady 
state of the two-qubit system. 
The striking differences between the predictions of each model become more evident 
if we consider the very strong coupling regime for the qubit-qubit interaction, e.g. 
$\lambda = 10 \Omega$, for which $\alpha_{+} \approx \alpha_{-} \approx 1/\sqrt{2}$,
i.e., the state $\rho_{\infty,m}$ becomes a maximally entangled state.
This may be illustrated if we calculate the concurrence corresponding
to the steady states; in Fig. (\ref{figure1}a) we have a plot of the concurrence 
relative to the steady state in equation (\ref{eq:asymptmic}), given by the microscopic model, 
as a function of time showing how the two-qubit state evolves to a maximally entangled state. On 
the other hand, as shown in Fig. (\ref{figure1}b), the state in equation (\ref{steadyphen}),
given by the phenomenological model is a stationary state having zero concurrence.

\begin{figure}
	\centering
		\includegraphics{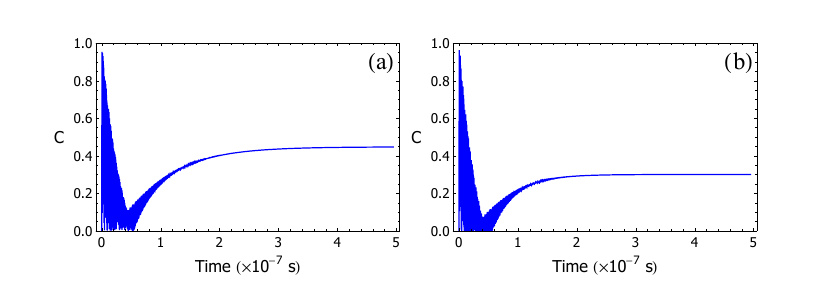}
	\caption{(Color online) Concurrence between qubits 1 and 2 as a function of time in the strong 
	coupling regime for an initial two qubit state $|\Psi\rangle_{q1,q2}=|1,0\rangle$ according to
	(a) the microscopic model; (b) the phenomenological model. 
	In both plots $\Omega = 4 \times 10^{9} s^{-1}$, $\lambda = \Omega$,
	$\gamma_0 = 0.001\times 5 \times 10^{10} s^{-1}$, $\Gamma = 5 \times 10^{10} s^{-1}$, 
	$\Omega_0 = 2\Omega$ and $T = 5 \times 10^{-4}$K.}
	\label{figure2}	
\end{figure}

\begin{figure}
	\centering
		\includegraphics{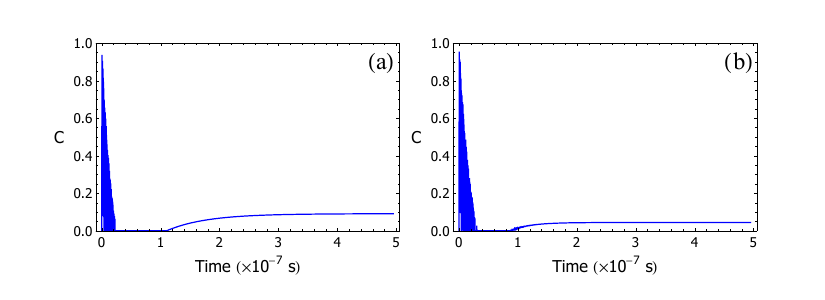}
	\caption{(Color online) Concurrence between qubits 1 and 2 as a function of time in the strong 
	coupling regime for an initial two qubit state $|\Psi\rangle_{q1,q2}=|1,0\rangle$ according to
	(a) the microscopic model; (b) the phenomenological model. 
	In both plots $\Omega = 4 \times 10^{9} s^{-1}$, $\lambda = \Omega$,
	$\gamma_0 = 0.001\times 5 \times 10^{10} s^{-1}$, $\Gamma = 5 \times 10^{10} s^{-1}$, 
	$\Omega_0 = 2\Omega$ and $T = 1.5 \times 10^{-2}$K.}
	\label{figure3}	
\end{figure}

Now we take $\lambda = \Omega$, i.e., strong coupling regime for the qubit-qubit interaction. 
In Fig. (\ref{figure2}a) we have the concurrence in the 
microscopic case, while in Fig. (\ref{figure2}b) we show the concurrence in the phenomenological case with initial conditions 
$|\psi(0)\rangle_{q1,q2} = |1,0\rangle$ for the two qubit system and a low temperature reservoir. We note an oscillatory 
pattern and a steady state value of entanglement (although with some differences) being attained in both cases. The phenomenon 
of stationary entanglement between the qubits has been already reported in the literature in similar systems 
\cite{sanchez11,paraoanu09} and it is confirmed here. We also note that for a higher temperature of the reservoir, as shown in 
Fig. (\ref{figure3}), the steady state values of entanglement decrease, as one would expect. 

We would like to remark that the dynamics of quantum entanglement according to the descriptions of both 
models may have a similar qualitative behaviour, as shown in Fig. (\ref{figure2}); e.g., both models predict stationary 
entanglement. However, the values of stationary concurrence may be significantly different. Thus, despite of the fact that some 
results arising from the phenomenological model may look as being physically acceptable, they should not be regarded as being 
correct. Entanglement has been shown to be a valuable resource for performing quantum information tasks. In reference \cite{kim00},
for instance, it is presented a protocol of teleportation of quantum entanglement, for which a ``critical value of minimum 
entanglement" is required. Thus, it would be interesting to know the amount of entanglement available at a given temperature of the
reservoir. We note that entanglement is underestimated according to the phenomenological model, compared to the
microscopic approach; e.g., at $T = 1.5\times 10^{-2}$ K [see Fig. (\ref{figure3})], the phenomenological model predicts a 
stationary value for the concurrence 51\% smaller than the microscopic model. For lower temperatures, $T = 5\times 10^{-4}$ K 
[see Fig. (\ref{figure2})], the relative difference is smaller, around 33\%. Thus, according to the more realistic microscopic model,
a fixed amount of entanglement (e.g., needed for a specific task), could be reached at a higher temperature.

\begin{figure}
	\centering
		\includegraphics{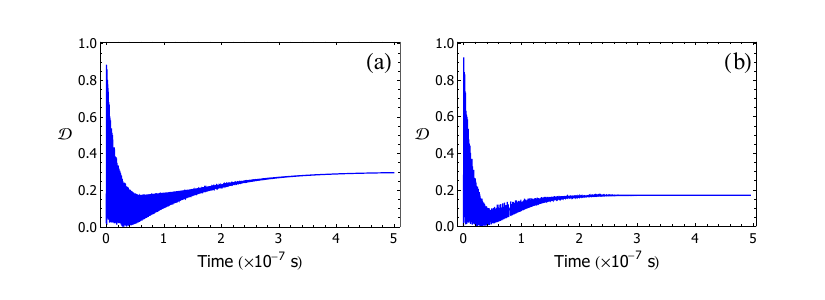}
	\caption{(Color online) Quantum discord between qubits 1 and 2 as a function of time in the strong 
	coupling regime for an initial two qubit state $|\Psi\rangle_{q1,q2}=|1,0\rangle$ according to
	(a) the microscopic model; (b) the phenomenological model. 
	In both plots $\Omega = 4 \times 10^{9} s^{-1}$, $\lambda = \Omega$,
	$\gamma_0 = 0.001\times 5 \times 10^{10} s^{-1}$, $\Gamma = 5 \times 10^{10} s^{-1}$, 
	$\Omega_0 = 2\Omega$ and $T = 5 \times 10^{-4}$K.}
	\label{figure4}	
\end{figure}

\begin{figure}
	\centering
		\includegraphics{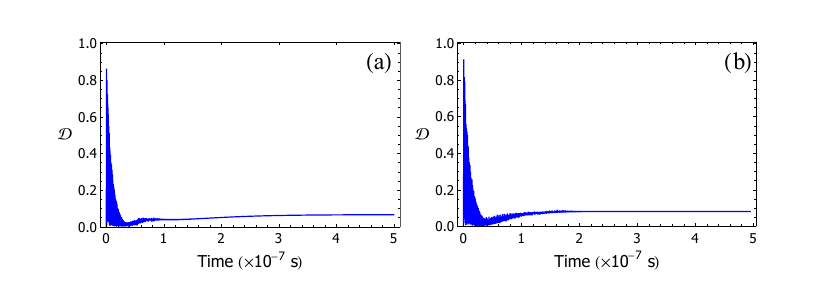}
	\caption{(Color online) Quantum discord between qubits 1 and 2 as a function of time in the strong 
	coupling regime for an initial two qubit state $|\Psi\rangle_{q1,q2}=|1,0\rangle$ according to
	(a) the microscopic model; (b) the phenomenological model. 
	In both plots $\Omega = 4 \times 10^{9} s^{-1}$, $\lambda = \Omega$,
	$\gamma_0 = 0.001\times 5 \times 10^{10} s^{-1}$, $\Gamma = 5 \times 10^{10} s^{-1}$, 
	$\Omega_0 = 2\Omega$ and $T = 1.5 \times 10^{-2}$K.}
	\label{figure5}	
\end{figure}

Discrepancies are also observed in the dynamics of the quantum discord, as seen in Fig. (\ref{figure4}). The discord reaches  
stationary values which are not the same according to each model. However, this may be qualitatively different from 
what we have found for the concurrence. We have that for lower temperatures [see Fig. (\ref{figure4})], the
steady state value of the discord is larger in the microscopic model, compared to the phenomenological model, but for higher temperatures
the situation is the opposite. For instance, if $T = 1.5\times 10^{-2}$ K, the discord [see Fig. (\ref{figure5})], differently from the 
concurrence [see Fig. (\ref{figure3})], is actually overestimated according to the phenomenological model. Here we have relative
differences of 42\% (below the correct value) at $T = 5\times 10^{-4}$ K, and 20\% (above the correct value) at $T = 1.5\times 10^{-2}$ K.

\begin{figure}
	\centering
		\includegraphics{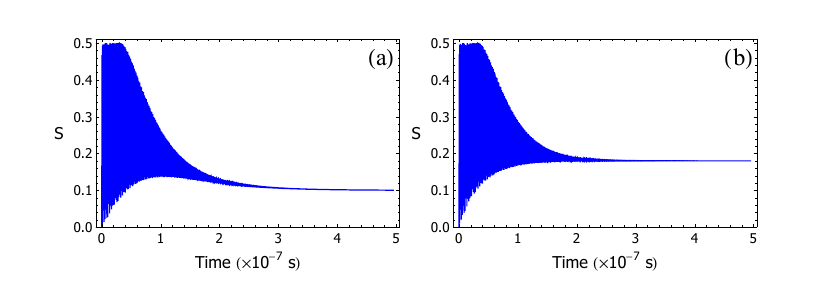}
	\caption{(Color online) Linear entropy of qubit 1 as a function of time in the strong coupling regime
	for an initial two qubit state $|\Psi\rangle_{q1,q2}=|1,0\rangle$ according to
	(a) the microscopic model; (b) the phenomenological model. 
	In both plots $\Omega = 4 \times 10^{9} s^{-1}$, $\lambda = \Omega$,
	$\gamma_0 = 0.001\times 5 \times 10^{10} s^{-1}$, $\Gamma = 5 \times 10^{10} s^{-1}$, 
	$\Omega_0 = 2\Omega$ and $T = 5 \times 10^{-4}$K.}
	\label{figure6}	
\end{figure}

\begin{figure}
	\centering
		\includegraphics{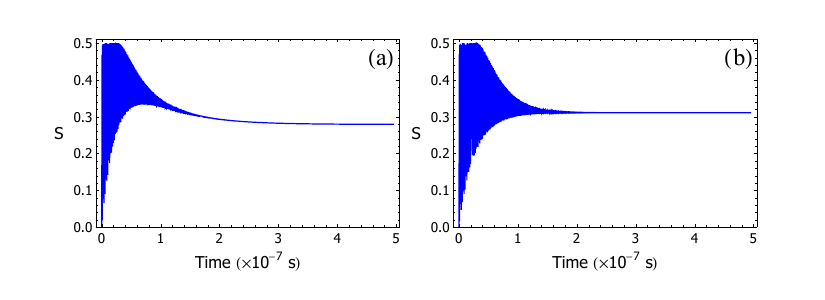}
	\caption{(Color online) Linear entropy of qubit 1 as a function of time in the strong coupling regime
	for an initial two qubit state $|\Psi\rangle_{q1,q2}=|1,0\rangle$ according to
	(a) the microscopic model; (b) the phenomenological model. 
	In both plots $\Omega = 4 \times 10^{9} s^{-1}$, $\lambda = \Omega$,
	$\gamma_0 = 0.001\times 5 \times 10^{10} s^{-1}$, $\Gamma = 5 \times 10^{10} s^{-1}$, 
	$\Omega_0 = 2\Omega$ and $T = 1.5 \times 10^{-2}$K.}
	\label{figure7}	
\end{figure}

Furthermore, we note that the evolution of the 
qubit 1 linear entropy is somehow associated to the concurrence, although it can not be used to quantify entanglement in case of 
a mixed global state. As shown in Fig. (\ref{figure6}), the linear entropy also attains steady state values according to both models. 
For higher temperatures of the reservoir (more thermal photons are injected into the system), the steady state value of the linear entropy 
of qubit 1 increases, which corresponds to a higher degree of mixedness of qubit 1, as shown in Fig.(\ref{figure7}).

\section{Comparison between microscopic and phenomenological models: weak coupling regime}

Now we turn our attention to the situation of very weak coupling regime, i.e. $\lambda \ll \Omega$. 
In this case the qubit-qubit interaction plays a less important role, and one would expect smaller 
differences between the results obtained from the two models. However, as we are going to see below,
the phenomenological approach may lead to incorrect results in this situation as well.

\subsection{Numerical results: weak coupling regime}

\begin{figure}
	\centering
		\includegraphics{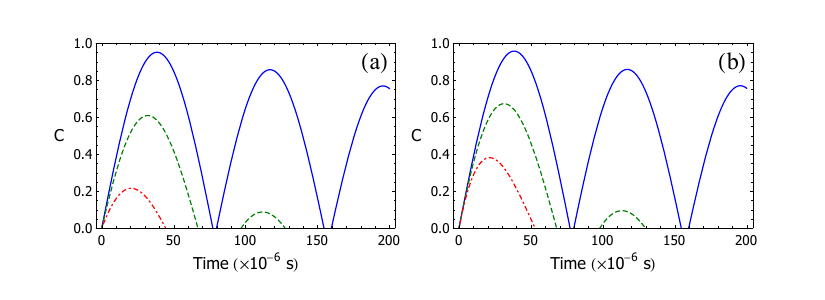}
	\caption{(Color online) Concurrence between qubits 1 and 2 as a function of time in the weak coupling regime
	for an initial two qubit state $|\Psi\rangle_{q1,q2}=|1,0\rangle$ according to
	(a) the microscopic model; (b) the phenomenological model. In both plots $\Omega = 5 \times 10^{6} s^{-1}$, 
	$\gamma_0 = 0.001\times 5 \times 10^{5} s^{-1}$, $\Gamma = 5 \times 10^{5} s^{-1}$,  $\lambda = 4 \times 10^{4} s^{-1}$, 
	and $\Omega_0 = 2\Omega$. The continuous (blue) curves correspond to a thermal bath at $T = 0.005$ K; the dashed (green) 
	curves to $T = 0.05$ K and the dot-dashed (red) curves to $T = 0.15$ K. (For interpretation of the references to color in 
	this figure caption, the reader is referred to the web version 
	of this paper.)}
	\label{figure8}	
\end{figure}

In Fig. (\ref{figure8}) we have plots of the concurrence; $C_m$ in Fig. (\ref{figure8}a) and $C_p$, in Fig. (\ref{figure8}b) 
as a function of time for an initial two-qubit state $|\Psi(0)\rangle_{q1,q2}=|1,0\rangle$.  
We note that for the reservoir at a temperature $T$ very close to zero, the concurrence curves show an 
oscillatory pattern as well as decay in both cases. As a matter of fact, despite the differences between the two models, 
the curves for $T \approx 0$ K virtually coincide. However, as the temperature of the reservoir is raised, we 
observe that the maxima of the concurrence are lower in the microscopic model compared to the curves obtained in the 
phenomenological model. We should point out that we have a typical pattern of entanglement sudden death in both cases, 
i.e., the concurrence vanishes for finite times.

\begin{figure}
	\centering
		\includegraphics{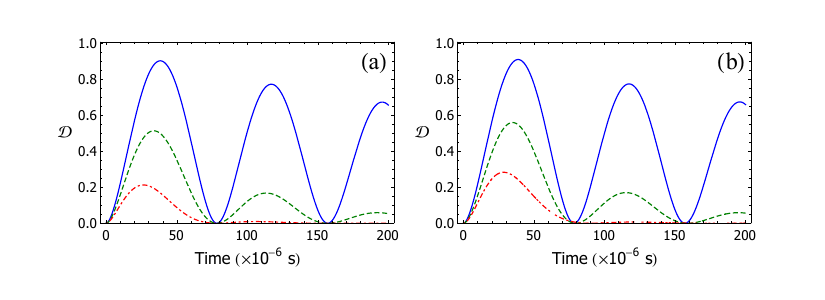}
	\caption{(Color online) Quantum discord between qubits 1 and 2 as a function of time in the weak coupling regime
	for an initial two qubit state $|\Psi\rangle_{q1,q2}=|1,0\rangle$ according to
	(a) the microscopic model; (b) the phenomenological model. In both plots $\Omega = 5 \times 10^{6} s^{-1}$, 
	$\gamma_0 = 0.001\times 5 \times 10^{5} s^{-1}$, $\Gamma = 5 \times 10^{5} s^{-1}$,  $\lambda = 4 \times 10^{4} s^{-1}$, 
	and $\Omega_0 = 2\Omega$. The continuous (blue) curves correspond to a thermal bath at $T = 0.005$ K; the dashed (green) 
	curves to $T = 0.05$ K and the dot-dashed (red) curves to $T = 0.15$ K. (For interpretation of the references to color in 
	this figure caption, the reader is referred to the web version 
	of this paper.)}
	\label{figure9}	
\end{figure}

A similar behaviour is observed for the quantum discord. The curves of this quantity are very close at 
$T\approx 0$ K but they become more discrepant for higher temperatures, as shown in Fig. (\ref{figure9}). 
However, differently from the concurrence, the quantum discord does not present sudden death, displaying
damped oscillations, instead \cite{werlang09}.

\begin{figure}
	\centering
		\includegraphics{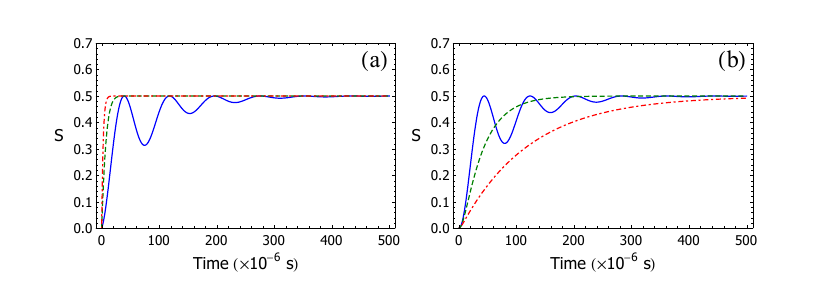}
	\caption{(Color online) Linear entropy of qubit 1 as a function of time for an initial two qubit state
	in the weak coupling regime $|\Psi\rangle_{q1,q2}=|1,0\rangle$ according to
	(a) the microscopic model; (b) the phenomenological model. In both plots $\Omega = 5 \times 10^{6} s^{-1}$, 
	$\gamma_0 = 0.01\times 5 \times 10^{5} s^{-1}$, $\Gamma = 5 \times 10^{5} s^{-1}$,  $\lambda = 4 \times 10^{4} s^{-1}$, and $\Omega_0 = 2\Omega$. 
	The continuous (blue) curves correspond to a thermal bath at $T = 0.005$ K; the dashed (green) curves to $T = 0.05$ K and the dot-dashed (red) 
	curves to $T = 0.15$ K. (For interpretation of the references to color in this figure caption, the reader is referred to the web version 
	of this paper.)}
	\label{figure10}	
\end{figure}

Regarding the linear entropy of qubit 1, though, we find that the microscopic and phenomenological models 
may yield contradictory time evolutions. 
In Fig. (\ref{figure10}) it is shown the linear entropy as a function of time for different temperatures 
of the reservoir. For a bath at $T \approx 0$ K, the linear entropy curves are virtually the same in both models: 
for the set of parameters chosen, they show oscillations and tend to the maximum value of $S_{max} = 0.5$, which
corresponds to a maximally mixed state. Nevertheless, the situation is very different if the reservoir is at finite temperature. 
Considering the microscopic model [see Fig. (\ref{figure10}a)], we notice that linear entropy increases  
at a faster rate if the temperature of the reservoir is higher. This is of course an intuitive and physically acceptable result, 
given that by increasing the temperature of the bath, a larger amount of noise is injected into the quantum system, and this should 
have a more destructive effect on the quantum coherence of qubit 1. On the other hand, according to the phenomenological model, 
the linear entropy increases at a faster rate if the temperature of the reservoir is lower, as shown in Fig. (\ref{figure10}b). 
This is in our opinion not realistic, as we would have coherent behaviour being induced in qubit 1 by a noisier bath. 
We note, though, that in both cases qubit 1 is eventually driven to a maximally mixed state, i.e., its linear entropy 
approaches the (equilibrium) expected asymptotic value of $S_{max} = 0.5$. 

\section{Conclusions}

We have made a comparison between two distinct models (microscopic $\times$ phenomenological) which describe the evolution 
of a system of two coupled two-level systems (qubits) in interaction with a thermal bath. We have studied the evolution
of quantities such as entanglement, quantum discord and the linear entropy relative to the two-qubit system.
We concluded that the results obtained from the {\it ad hoc} (phenomenological) model are in general not 
in accord with the ones obtained from the microscopic model, in both the strong and weak (qubit-qubit) coupling regimes. 

Firstly we have analyzed the case of strong coupling regime, for which we expect a more significant disagreement 
between the results from each model. We have shown analytically that according 
to the phenomenological model, the two-qubit system evolves to a steady state whose corresponding density 
operator is not a thermal equilibrium state, while the microscopic model gives the correct prediction. 
Moreover, in the strong coupling regime (for $T \neq 0$ K), the qualitative 
behaviour of entanglement of the system is somewhat similar in the framework of both models, but the
steady state values may be considerably different, as shown in Fig. (\ref{figure2}). This should 
be relevant for the implementation of quantum information tasks requiring a minimum amount of entanglement, 
as discussed in \cite{kim00}. Besides, the values of the steady states of both the quantum discord and the linear 
entropy of qubit 1 are also different for each model. 

Yet, by assuming a weak coupling between the qubits, one could expect the predictions from both models to be in 
better agreement with each other. This is true if $T = 0$ K, but we have found important differences 
if the reservoir is at finite temperature. Concerning the entanglement between the two qubits, the differences are 
small although they are more noticeable at higher temperatures; the microscopic model predicts a more destructive action 
of the thermal noise compared to the phenomenological construct, as we note in Fig. (\ref{figure8}). This may be readily
understood because contrarily to what happens $T = 0$ K, if the bath is at finite temperature, photons are injected from 
the reservoir into the system. As a consequence, in the realm of the microscopic model, those thermal photons will be 
inducing transitions between the dressed levels of the two qubit system, and we expect a more disordered 
evolution for higher temperatures of the reservoir. Interestingly, the discrepancies are more evident if 
one focuses on the evolution of the state purity of qubit 1 despite being weakly coupled to qubit 2. Even though the 
curves of the linear entropy (state purity) are the same (according to each model), if the bath is at 
$T = 0$ K, for higher temperatures the phenomenological and microscopic models lead to conflicting results. While 
according to the microscopic model qubit 1 evolves more rapidly to a mixed state for higher temperatures 
of the reservoir, the phenomenological model predicts the opposite behaviour, as seen in Fig. (\ref{figure10}). 

Thus, we have demonstrated here that oversimplified phenomenological models used to describe the evolution of a two-qubit 
system asymmetrically coupled to an environment, may lead to misleading results even in the weak (qubit-qubit) coupling regime, 
and therefore there is need of a more appropriate modeling procedure.
We should point out that, as microscopic master equations may be hard to construct, perturbative methods as discussed in 
\cite{volovich15} could be very useful to treat such quantum open composite systems. 

\section*{Acknowledgements}

G.L.D. would like to thank CAPES (Coordena\c c\~ao de Aperfei\c coamento 
de Pessoal de N\'\i vel Superior) under grant 2011/899872, 
for financial support. This work was also supported by CNPq 
(Conselho Nacional para o Desenvolvimento Cient\'\i fico e Tecnol\'ogico) 
and FAPESP (Funda\c c\~ao de Amparo \`a Pesquisa do Estado de S\~ao Paulo), 
through the INCT-IQ (National Institute for Science and Technology of Quantum Information) 
under grant 2008/57856-6 and the CePOF (Optics and Photonics 
Research Center) under grant 2005/51689-2, Brazil.

\appendix

\section{Equations for the matrix elements and their solutions in the strong coupling regime - microscopic case}

From the master equation (\ref{mestramic}) we may obtain a set of coupled differential equations for the dressed
state populations of the two-qubit system. Populations:

\begin{eqnarray}
\dot{\rho}_{aa}(t) & = & -\left(\bar{c}_{I}+\bar{c}_{II}\right)\rho_{aa}(t)+c_{I}\rho_{bb}(t)+c_{II}\rho_{cc}(t)\,,\nonumber \\
\nonumber \\
\dot{\rho}_{bb}(t) & = & \bar{c}_{I}\rho_{aa}(t)-\left(c_{I}+\bar{c}_{II}\right)\rho_{bb}(t)+c_{II}\rho_{dd}(t)\,,\nonumber \\
\label{eq:pop1}\\
\dot{\rho}_{cc}(t) & = & \bar{c}_{II}\rho_{aa}(t)-\left(\bar{c}_{I}+c_{II}\right)\rho_{cc}(t)+c_{I}\rho_{dd}(t)\,,\nonumber \\
\nonumber \\
\dot{\rho}_{dd}(t) & = & \bar{c}_{II}\rho_{bb}(t)+\bar{c}_{I}\rho_{cc}(t)-\left(c_{I}+c_{II}\right)\rho_{dd}(t)\,,\nonumber 
\end{eqnarray}
and coherences

\begin{eqnarray}
\dot{\rho}_{ab}(t) & = & \left[i\left(\frac{\sqrt{\lambda^{2}+4\Omega^{2}}-\lambda}{2}\right)-\frac{\left(c_{I}+\bar{c}_{I}+
2\bar{c}_{II}\right)}{2}\right]\rho_{ab}(t)+c_{II}\rho_{cd}(t)\,,\nonumber\\
\nonumber\\
\dot{\rho}_{ac}(t) & = & \left[i\left(\frac{\sqrt{\lambda^{2}+4\Omega^{2}}+\lambda}{2}\right)-\frac{\left(2\bar{c}_{I}+
c_{II}+\bar{c}_{II}\right)}{2}\right]\rho_{ac}(t)-c_{I}\rho_{bd}(t)\,,\nonumber\\
\nonumber\\
\dot{\rho}_{ad}(t) & = & \left[i\,\sqrt{\lambda^{2}+4\Omega^{2}}-\frac{\left(c_{I}+c_{II}+\bar{c}_{I}+
\bar{c}_{II}\right)}{2}\right]\rho_{ad}(t)\,,\nonumber\\
\label{eq:coher1} \\
\dot{\rho}_{bc}(t) & = & \left[i\lambda-\frac{\left(c_{I}+c_{II}+\bar{c}_{I}+\bar{c}_{II}\right)}{2}\right]\rho_{bc}(t)\,,\nonumber\\
\nonumber\\
\dot{\rho}_{bd}(t) & = & \left[i\left(\frac{\sqrt{\lambda^{2}+4\Omega^{2}}+\lambda}{2}\right)-\frac{\left(2c_{I}+c_{II}+
\bar{c}_{II}\right)}{2}\right]\rho_{bd}(t)-\bar{c}_{I}\rho_{ac}(t)\,,\nonumber\\
\nonumber\\
\dot{\rho}_{cd}(t) & = & \left[i\left(\frac{\sqrt{\lambda^{2}+4\Omega^{2}}-\lambda}{2}\right)-\frac{\left(c_{I}+2c_{II}+
\bar{c}_{I}\right)}{2}\right]\rho_{cd}(t)+\bar{c}_{II}\rho_{ab}(t)\,.\nonumber
\end{eqnarray}
The corresponding solutions are:

\begin{eqnarray}
k\,\rho_{aa}\left(t\right) & = & c_{I}\, c_{II}+e^{-\left(c_{I}+\overline{c}_{I}\right)t}\left\{ \overline{c}_{I}\, c_{II}[\rho_{aa}\left(0\right)+\rho_{cc}\left(0\right)]-c_{I}\, c_{II}[\rho_{bb}\left(0\right)+\rho_{dd}\left(0\right)]\right\} \nonumber\\
\nonumber\\
 &  & +e^{-\left(c_{II}+\overline{c}_{II}\right)t}\left\{ c_{I}\,\overline{c}_{II}[\rho_{aa}\left(0\right)+\rho_{bb}\left(0\right)]-c_{I}\, c_{II}[\rho_{cc}\left(0\right)+\rho_{dd}\left(0\right)]\right\} \nonumber\\
\nonumber\\
 &  & +e^{-\left(c_{I}+c_{II}+\overline{c}_{I}+\overline{c}_{II}\right)t}\left\{ \overline{c}_{I}\,\overline{c}_{II}\rho_{aa}\left(0\right)-c_{I}\,\overline{c}_{II}\rho_{bb}\left(0\right)-\overline{c}_{I}\, c_{II}\rho_{cc}\left(0\right)+c_{I}\, c_{II}\rho_{dd}\left(0\right)\right\} \,,\nonumber
\end{eqnarray}

\begin{eqnarray}
k\,\rho_{bb}\left(t\right) & = & \overline{c}_{I}\, c_{II}+e^{-\left(c_{I}+\overline{c}_{I}\right)t}\left\{ -\overline{c}_{I}\, c_{II}[\rho_{aa}\left(0\right)+\rho_{cc}\left(0\right)]+c_{I}\, c_{II}[\rho_{bb}\left(0\right)+\rho_{dd}\left(0\right)]\right\} \nonumber\\
\nonumber\\
 &  & +e^{-\left(c_{II}+\overline{c}_{II}\right)t}\left\{ \overline{c}_{I}\,\overline{c}_{II}[\rho_{aa}\left(0\right)+\rho_{bb}\left(0\right)]-\overline{c}_{I}\, c_{II}[\rho_{cc}\left(0\right)+\rho_{dd}\left(0\right)]\right\} \nonumber\\
\label{eq:solpop1}\\
 &  & +e^{-\left(c_{I}+c_{II}+\overline{c}_{I}+\overline{c}_{II}\right)t}\left\{ -\overline{c}_{I}\,\overline{c}_{II}\rho_{aa}\left(0\right)+c_{I}\,\overline{c}_{II}\rho_{bb}\left(0\right)+\overline{c}_{I}\, c_{II}\rho_{cc}\left(0\right)-c_{I}\, c_{II}\rho_{dd}\left(0\right)\right\} \,,
\nonumber
\end{eqnarray}

\begin{eqnarray*}
k\,\rho_{cc}\left(t\right) & = & c_{I}\,\overline{c}_{II}+e^{-\left(c_{I}+\overline{c}_{I}\right)t}\left\{ \overline{c}_{I}\,\overline{c}_{II}[\rho_{aa}\left(0\right)+\rho_{cc}\left(0\right)]-c_{I}\,\overline{c}_{II}[\rho_{bb}\left(0\right)+\rho_{dd}\left(0\right)]\right\}\\
\\
 &  & +e^{-\left(c_{II}+\overline{c}_{II}\right)t}\left\{ -c_{I}\,\overline{c}_{II}[\rho_{aa}\left(0\right)+\rho_{bb}\left(0\right)]+c_{I}\, c_{II}[\rho_{cc}\left(0\right)+\rho_{dd}\left(0\right)]\right\} \nonumber\\
\\
 &  & +e^{-\left(c_{I}+c_{II}+\overline{c}_{I}+\overline{c}_{II}\right)t}\left\{ -\overline{c}_{I}\,\overline{c}_{II}\rho_{aa}\left(0\right)+c_{I}\,\overline{c}_{II}\rho_{bb}\left(0\right)+\overline{c}_{I}\, c_{II}\rho_{cc}\left(0\right)-c_{I}\, c_{II}\rho_{dd}\left(0\right)\right\} \,,
\end{eqnarray*}

\begin{eqnarray*}
k\,\rho_{dd}\left(t\right) & = & \overline{c}_{I}\,\overline{c}_{II}+e^{-\left(c_{I}+\overline{c}_{I}\right)t}\left\{ -\overline{c}_{I}\,\overline{c}_{II}[\rho_{aa}\left(0\right)+\rho_{cc}\left(0\right)]+c_{I}\,\overline{c}_{II}[\rho_{bb}\left(0\right)+\rho_{dd}\left(0\right)]\right\}\\
\\
 &  & +e^{-\left(c_{II}+\overline{c}_{II}\right)t}\left\{ -\overline{c}_{I}\,\overline{c}_{II}[\rho_{aa}\left(0\right)+\rho_{bb}\left(0\right)]+\overline{c}_{I}\, c_{II}[\rho_{cc}\left(0\right)+\rho_{dd}\left(0\right)]\right\} \\
\\
 &  & +e^{-\left(c_{I}+c_{II}+\overline{c}_{I}+\overline{c}_{II}\right)t}\left\{ \overline{c}_{I}\,\overline{c}_{II}\rho_{aa}\left(0\right)-c_{I}\,\overline{c}_{II}\rho_{bb}\left(0\right)-\overline{c}_{I}\, c_{II}\rho_{cc}\left(0\right)+c_{I}\, c_{II}\rho_{dd}\left(0\right)\right\} \,,
\end{eqnarray*}
where $k=\left(c_{I}+\overline{c}_{I}\right)\left(c_{II}+\overline{c}_{II}\right)$ and the coefficients $c_i$ are defined in (\ref{coeffs}).

And the coherences are:

\[
\rho_{ab}\left(t\right)=\frac{e^{\left[i\left(\frac{\sqrt{\lambda^{2}+4\Omega^{2}}-\lambda}{2}\right)-\left(\frac{c_{I}+\overline{c}_{I}}{2}\right)\right]t}}{c_{II}+\overline{c}_{II}}\left\{ \left[c_{II}+e^{-\left(c_{II}+\overline{c}_{II}\right)t}\,\overline{c}_{II}\right]\rho_{ab}\left(0\right)+\left[1-e^{-\left(c_{II}+\overline{c}_{II}\right)t}\right]c_{II}\,\rho_{cd}\left(0\right)\right\} \,,
\]

\[
\rho_{ac}\left(t\right)=\frac{e^{\left[i\left(\frac{\sqrt{\lambda^{2}+4\Omega^{2}}+\lambda}{2}\right)-\left(\frac{c_{II}+\overline{c}_{II}}{2}\right)\right]t}}{c_{I}+\overline{c}_{I}}\left\{\left[c_{I}+e^{-\left(c_{I}+\overline{c}_{I}\right)t}\,\overline{c}_{I}\right]\rho_{ac}\left(0\right)-\left[1-e^{-\left(c_{I}+\overline{c}_{I}\right)t}\right]c_{I}\,\rho_{bd}\left(0\right)\right\} \,,
\]

\[
\rho_{ad}\left(t\right)=e^{-\left(\frac{c_{I}+c_{II}+\overline{c}_{I}+\overline{c}_{II}}{2}-i\,\sqrt{\lambda^{2}+4\Omega^{2}}\right)t}\rho_{ad}\left(0\right)\,,
\]

\[
\rho_{bc}\left(t\right)=e^{-\left(\frac{c_{I}+c_{II}+\overline{c}_{I}+\overline{c}_{II}}{2}-i\lambda\right)t}\rho_{bc}\left(0\right)\,,
\]

\[
\rho_{bd}\left(t\right)=\frac{e^{\left[i\left(\frac{\sqrt{\lambda^{2}+4\Omega^{2}}+\lambda}{2}\right)-\left(\frac{c_{II}+\overline{c}_{II}}{2}\right)\right]t}}{c_{I}+\overline{c}_{I}}\left\{ -\left[1-e^{-\left(c_{I}+\overline{c}_{I}\right)t}\right]\overline{c}_{I}\,\rho_{ac}\left(0\right)+\left[\overline{c}_{I}+e^{-\left(c_{I}+\overline{c}_{I}\right)t}c_{I}\right]\rho_{bd}\left(0\right)\right\} \,,
\]

\[
\rho_{cd}\left(t\right)=\frac{e^{\left[i\left(\frac{\sqrt{\lambda^{2}+4\Omega^{2}}-\lambda}{2}\right)-\left(\frac{c_{I}+\overline{c}_{I}}{2}\right)\right]t}}{c_{II}+\overline{c}_{II}}\left\{ \left[1-e^{-\left(c_{II}+\overline{c}_{II}\right)t}\right]\overline{c}_{II}\,\rho_{ab}\left(0\right)+\left[\overline{c}_{II}+e^{-\left(c_{II}+\overline{c}_{II}\right)t}c_{II}\right]\rho_{cd}\left(0\right)\right\} \,.
\]

\section{Equations for the matrix elements in the strong coupling regime - phenomenological case}

Set of coupled differential equations for the matrix elements of the two-qubit system in the strong coupling
regime obtained from the phenomenological master equation. Populations:

\begin{eqnarray}
\dot{\rho}_{11}\left(t\right) & = & -\overline{\gamma}\rho_{11}\left(t\right)+\gamma\rho_{22}\left(t\right)+\frac{i\lambda}{2}\rho_{14}\left(t\right)-\frac{i\lambda}{2}\rho_{41}\left(t\right)\nonumber\\
\nonumber\\
\dot{\rho}_{22}\left(t\right) & = & \overline{\gamma}\rho_{11}\left(t\right)-\gamma\rho_{22}\left(t\right)+\frac{i\lambda}{2}\rho_{23}\left(t\right)-\frac{i\lambda}{2}\rho_{32}\left(t\right)\nonumber\\
\label{eq:popphen}\\
\dot{\rho}_{33}\left(t\right) & = & -\overline{\gamma}\rho_{33}\left(t\right)+\gamma\rho_{44}\left(t\right)+\frac{i\lambda}{2}\rho_{32}\left(t\right)-\frac{i\lambda}{2}\rho_{23}\left(t\right)\nonumber\\
\nonumber\\
\dot{\rho}_{44}\left(t\right) & = & \overline{\gamma}\rho_{33}\left(t\right)-\gamma\rho_{44}\left(t\right)+\frac{i\lambda}{2}\rho_{41}\left(t\right)-\frac{i\lambda}{2}\rho_{14}\left(t\right),\nonumber
\end{eqnarray}

and for the coherences,

\begin{eqnarray}
\dot{\rho}_{12}\left(t\right) & = & \left[i\Omega-\frac{\left(\gamma+\overline{\gamma}\right)}{2}\right]\rho_{12}\left(t\right)+\frac{i\lambda}{2}\rho_{13}\left(t\right)-\frac{i\lambda}{2}\rho_{42}\left(t\right)\nonumber\\
\nonumber\\
\dot{\rho}_{13}\left(t\right) & = & \frac{i\lambda}{2}\rho_{12}\left(t\right)+\left(i\Omega-\overline{\gamma}\right)\rho_{13}\left(t\right)+\gamma\rho_{24}\left(t\right)-\frac{i\lambda}{2}\rho_{43}\left(t\right)\nonumber\\
\nonumber\\
\dot{\rho}_{14}\left(t\right) & = & \left[2i\Omega-\frac{\left(\gamma+\overline{\gamma}\right)}{2}\right]\rho_{14}\left(t\right)+\frac{i\lambda}{2}\rho_{11}\left(t\right)-\frac{i\lambda}{2}\rho_{44}\left(t\right)\nonumber\\
\label{eq:cohephen}\\
\dot{\rho}_{23}\left(t\right) & = & \frac{i\lambda}{2}\rho_{22}\left(t\right)-\frac{\left(\gamma+\overline{\gamma}\right)}{2}\rho_{23}\left(t\right)-\frac{i\lambda}{2}\rho_{33}\left(t\right)\nonumber\\
\nonumber\\
\dot{\rho}_{24}\left(t\right) & = & -\frac{i\lambda}{2}\rho_{34}\left(t\right)+\left(i\Omega-\gamma\right)\rho_{24}\left(t\right)+\overline{\gamma}\rho_{13}\left(t\right)+\frac{i\lambda}{2}\rho_{21}\left(t\right)\nonumber\\
\nonumber\\
\dot{\rho}_{34}\left(t\right) & = & \left[i\Omega-\frac{\left(\gamma+\overline{\gamma}\right)}{2}\right]\rho_{34}\left(t\right)-\frac{i\lambda}{2}\rho_{24}\left(t\right)+\frac{i\lambda}{2}\rho_{13}\left(t\right).\nonumber
\end{eqnarray}



\end{document}